\documentclass{elsart}%
\usepackage{amsfonts}
\usepackage{amsmath}%
\usepackage{amssymb}%
\usepackage{graphicx}
\begin{document}
%
\begin{frontmatter}%
%

\title
{Profit Maximization, Industry Structure, and Competition: A critique of neoclassical theory}%
%

\author{Steve Keen (University of Western Sydney)}%
%

\collab{Russell Standish (University of New South Wales)}%
%

\address{Locked Bag 1797, Penrith 1797, Australia. s.keen@uws.edu.au}%
%

\begin{abstract}
Neoclassical economics has two theories of competition between
profit-maximizing firms (Marshallian and Cournot-Nash) that start from
different premises about the degree of strategic interaction between firms,
yet reach the same result, that market price falls as the number of firms in
an industry increases. The Marshallian argument is strictly false. We
integrate the different premises, and establish that the optimal level of
strategic interaction between competing firms is zero. Simulations support our
analysis and reveal intriguing emergent behaviors.

\end{abstract}%
%

%

\end{frontmatter}%

\section{A popular fallacy}

The underlying assumption of the Marshallian model is that the $i^{th}$\ firm
in a competitive industry does not react strategically to the hypothetical
actions of other firms. In an $n$ firm industry where the output of the
$i^{th}$ firm is $q_{i}$, this assumption, known as \textquotedblleft
atomism\textquotedblright\ or \textquotedblleft price-taking\textquotedblright%
,\cite[pp. 314, 383]{MasColell} means that $\frac{\partial q_{i}}{\partial
q_{j}}=0$\ $\forall i\neq j$. This model then claims that the market demand
function $P\left(  Q\right)  $ (where $Q=\sum\limits_{i=1}^{n}q_{i}$) has the
dual properties that $P^{\prime}\left(  Q\right)  <0$ and $P^{\prime}\left(
q_{i}\right)  =0$ for large $n$. Elementary calculus shows this is false:%

\begin{equation}
\dfrac{dP}{dq_{i}}=\dfrac{dP}{dQ}\dfrac{dQ}{dq_{i}}\label{StiglerRelation}%
\end{equation}

where $\dfrac{dQ}{dq_{i}}=1$ given $\frac{\partial q_{i}}{\partial q_{j}}%
=0$\ $\forall i\neq j$. Therefore $\dfrac{dP}{dq_{i}}=\dfrac{dP}{dQ}$
(\cite{Stigler}).

This false belief is essential to the Marshallian derivation of the model of
\textquotedblleft perfect competition\textquotedblright, which occurs when
$P\left(  Q\right)  $---the marginal benefit to society---equals the marginal
cost of production $MC\left(  Q\right)  $.\cite[p. 322]{MasColell} The
derivation starts from the proposition that a profit maximizing firm will
produce where its Marginal Cost $\left(  \frac{d}{dq_{i}}TC\left(
q_{i}\right)  \right)  $ equals its Marginal\ Revenue $\left(  \frac{d}%
{dq_{i}}\left(  P\left(  Q\right)  q_{i}\right)  \right)  $:%

\begin{equation}
\pi_{neo}:\frac{d}{dq_{i}}\pi\left(  q_{i}\right)  =MR\left(  q_{i}\right)
-MC\left(  q_{i}\right)  =0\label{ProfitMaxCournot}%
\end{equation}

\bigskip Given this alleged profit-maximization rule and the \textquotedblleft
assumption\textquotedblright\ that $P^{\prime}\left(  q_{i}\right)  =0$, it
followed that for perfect competition, price equalled marginal cost. Since the
assumption is logically incompatible with $P^{\prime}\left(  Q\right)  <0$,
the Marshallian derivation of perfect competition fails.

The Cournot-Nash model is not dependent on this fallacy, arguing instead that
strategic interactions lead to a Nash equilibrium in which market price
converges to marginal cost as the number of firms increases (\cite{Vega};
\cite[pp. 411-413.]{MasColell})---though this process is, at best, highly
conditional at best (see \cite[pp. 417-423]{MasColell}). Standard neoclassical
analysis assumes that firms \textit{will} strategically interact, and
calculates the $i^{th}$ firm's best response on this basis under various
conditions. We instead treat $\frac{\partial q_{i}}{\partial q_{j}}$, the
response of the $i^{th}$ firm to a hypothetical change in output by the
$j^{th}$, as a decision variable, and consider what is its optimal value of
$\frac{\partial q_{i}}{\partial q_{j}}$ for a profit-maximizing firm. As a
preliminary, we show that the proposition that (\ref{ProfitMaxCournot})
maximizes profits for the $i^{th}$ firm is false.

\section{The true profit maximization formula}

In a multi-firm industry, the profit maximum is given by the zero, not of its
\emph{partial} derivative, but its \emph{total} derivative---since the actions
of other firms affect the profitability of any given firm, even though (or
rather, especially because) the $i^{th}$ firm cannot control what the other
firms in the industry do. Maximizing profit while ignoring what other firms do
is rather like rowing a boat to a specific location while ignoring the wind
and tides. The profit maximum for the $i^{th}$ firm is therefore given by:%

\begin{equation}
\dfrac{d}{dQ}\pi\left(  q_{i}\right)  =\dfrac{d}{dQ}\left(  P\left(  Q\right)
q_{i}-TC\left(  q_{i}\right)  \right)  =0\label{ProfitMaxCondition}%
\end{equation}
\hspace{5mm}

Since $Q=\sum\limits_{j=1}^{n}q_{j}$, (\ref{ProfitMaxCondition}) can be
expanded to%

\begin{equation}
\sum\limits_{j=1}^{n}\left(  \dfrac{\partial}{\partial q_{j}}\left(  P\left(
\sum\limits_{j=1}^{n}q_{j}\right)  q_{i}-TC\left(  q_{i}\right)  \right)
\dfrac{dq_{j}}{dQ}\right)  =0\label{ProfitMaxConditionExpanded}%
\end{equation}
\hspace{5mm}

Converting (\ref{ProfitMaxConditionExpanded}) into an expression in terms of
reaction coefficients $\frac{\partial q_{i}}{\partial q_{j}}$ yields:%

\begin{equation}
P\sum_{j=1}^{n}\left(  \frac{\partial q_{i}}{\partial q_{j}}\sum_{k=1}%
^{n}\frac{\partial q_{j}}{\partial q_{k}}\right)  +q_{i}\frac{dP}{dQ}%
\sum_{j=1}^{n}\sum_{k=1}^{n}\frac{\partial q_{j}}{\partial q_{k}}-MC\left(
q_{i}\right)  \sum_{j=1}^{n}\frac{\partial q_{i}}{\partial q_{j}%
}=0\label{GeneralProfitMaximum}%
\end{equation}
With the Marshallian assumption of \textquotedblleft atomism\textquotedblright%
, $\frac{\partial q_{i}}{\partial q_{j}}=0$ $\forall i\neq j$, and equation
(\ref{GeneralProfitMaximum}) reduces to%

\begin{equation}
P+nq_{i}\frac{dP}{dQ}-MC\left(  q_{i}\right)  =0\label{ProfitMaximumThetaZero}%
\end{equation}

\bigskip This contradicts the neoclassical belief, epitomized by
(\ref{ProfitMaxCournot}),\ that, in the context of atomistic behavior, profit
is maximized by equating marginal revenue and marginal cost.
(\ref{ProfitMaximumThetaZero})\ can be rearranged to yield:%

\begin{equation}
MR\left(  q_{i}\right)  -MC\left(  q_{i}\right)  =\frac{n-1}{n}\left(
P-MC\left(  q_{i}\right)  \right) \label{MR_MCGapRule}%
\end{equation}

\bigskip This equals zero only in the case of a monopoly---which is the one
time that the accepted Marshallian formula is correct. At all other times, the
profit maximum for an individual firm will occur where \textquotedblleft
marginal revenue\textquotedblright\ \emph{exceeds} \textquotedblleft marginal
cost\textquotedblright. As a consequence, the Marshallian model leads to
industry output being independent of the number of firms in it.

In Cournot-Nash game theoretic analysis, firms decide their own actions on the
basis of the expected reactions of other firms, in such a way that each firm's
\textquotedblleft best response\textquotedblright\ is to set $MR\left(
q_{i}\right)  =MC\left(  q_{i}\right)  $. In our terms, this is equivalent to
setting $\frac{\partial q_{i}}{\partial q_{j}}=\dfrac{1}{nE}$---where E is the
market elasticity of demand ($E=\dfrac{P}{Q}\dfrac{dQ}{dP}$). Our equation
(\ref{GeneralProfitMaximum}) lets us combine Marshallian and Cournot-Nash
analysis, by treating $\frac{\partial q_{i}}{\partial q_{j}}$ as a decision
variable whose optimum value can be identified by the firm. In this paper, we
consider an industry of $n$ identical firms (a common heuristic device in
economic theory) so that $\frac{\partial q_{i}}{\partial q_{j}}=\theta
$\ $\forall i\neq j$ and $\frac{\partial q_{i}}{\partial q_{i}}=1$, where
$\theta$\ can take on any value. Then (\ref{GeneralProfitMaximum}) reduces to:%

\begin{equation}
\left(  n-1\right)  P\theta+P+nq_{i}\dfrac{dP}{dQ}=MC\left(  q_{i}\right)
\label{ProfitMaximumTheta}%
\end{equation}
\hspace{5mm}

This defines the maximum profit achievable by the individual firm in the
context of strategic behavior---if each firm reacts to output changes by other
firms with a reaction coefficient of $\theta$. We can now consider what value
of $\theta$\ would be chosen by a profit-maximizing firm. It transpires that
the optimum value of this parameter is in fact zero.

\section{True profit-maximizing behavior}

The optimum value for $\theta$ for the $i^{th}$ firm occurs where $\frac
{d}{d\theta}\pi\left(  q_{i}\right)  =0$. This condition reduces to:%

\begin{equation}
\frac{d}{d\theta}\pi\left(  q_{i}\right)  =\frac{1}{n}\frac{d}{d\theta
}Q\left(  P+nq_{i}\frac{dP}{dQ}-MC\left(  q_{i}\right)  \right)
\label{ProfitMaxwrtTheta}%
\end{equation}

Since it can be shown that $\frac{d}{d\theta}Q\neq0$, (\ref{ProfitMaxwrtTheta}%
) equals zero iff $P+nq_{i}\frac{dP}{dQ}-MC\left(  q_{i}\right)  =0$.\ As
established above, this requires that $\theta=0$. Firms thus achieve higher
profits if they \emph{do not} react strategically to each other. In the
classic words of the movie \textit{War\ Games}, firms may conclude that
Cournot-Nash strategic interaction is \textquotedblleft A curious game.\ The
only winning strategy is not to play.\textquotedblright\ We consider this
question using a multi-agent model of instrumentally rational profit
maximizers facing comparable marginal cost functions.

\section{ Operationally rational profit-maximizers}

Our hypothetical market has a linear demand curve ($P=a-bQ$) and a given
number $n$ of profit-maximizing agents. Firm $i$ chooses an initial output
level $q_{i,0}$ and a fixed amount by which to vary output at each step
$\delta q_{i}$. If profit falls after an iteration, $i$ reverses the sign of
$\delta q_{i}$ for the next iteration.\footnote{The programs for this paper
can be downloaded from $\emph{www.debunking-economics.com/totf}$.} Total cost
functions for the firms are identical:%

\begin{equation}
tc\left(  q,n\right)  =k+Cq+\dfrac{1}{2}Dnq^{2}+\dfrac{1}{3}En^{2}%
q^{3}\label{TotalCost}%
\end{equation}

\begin{figure}
[ptbh]
\begin{center}
\includegraphics[
height=2.8573in,
width=3.7913in
]%
{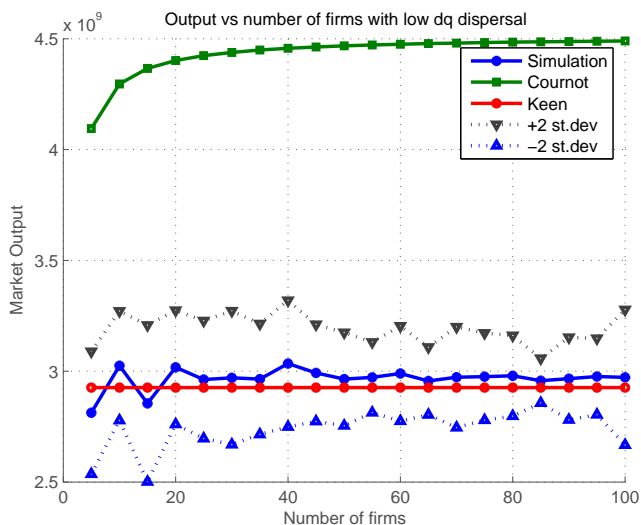}%
\caption{Outcome of Monte Carlo simulations with low dispersal of $\delta q$
(2\% Cournot)}%
\label{RisingMC_Low_dq}%
\end{center}
\end{figure}

In the following simulations, $a=800$ and $b=10^{-7}$, $k=10^{6}$, $C=10$,
$D=10^{-8}$and $E=10^{-17}$ and $n$ ranges between $5$ and $100$ (higher
values made no significant difference to our results.); the fixed $\delta
q_{i}$ for each firm is drawn from a normal distribution $N\left(
m,\sigma\right)  $ where $m=0$ and $\sigma$ is a given fraction of
the\ Cournot prediction. Monte Carlo simulations reveal a rich range of
interactions, and in general show that instrumentally rational
profit-maximizers will learn \textquotedblleft not to play\textquotedblright%
\ the Cournot-Nash game. For low $\sigma$, output converges to the
\textquotedblleft Keen\textquotedblright\ equilibrium given by
(\ref{ProfitMaximumThetaZero})\ for all values of $n$ (Figure
\ref{RisingMC_Low_dq}).%

\begin{figure}
[ptbh]
\begin{center}
\includegraphics[
height=2.8573in,
width=3.7913in
]%
{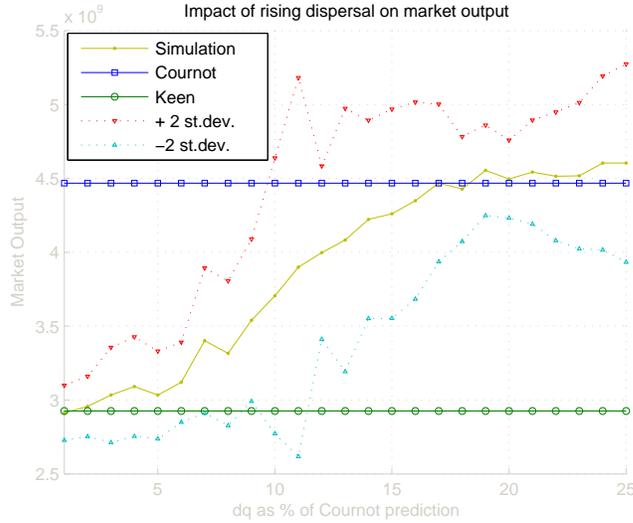}%
\caption{Convergence to\ Cournot-Nash prediction as $\delta q$ rises with
constant $n=50$}%
\label{DispersalRisingMC}%
\end{center}
\end{figure}

\begin{figure}
[ptb]
\begin{center}
\includegraphics[
height=2.5365in,
width=3.7913in
]%
{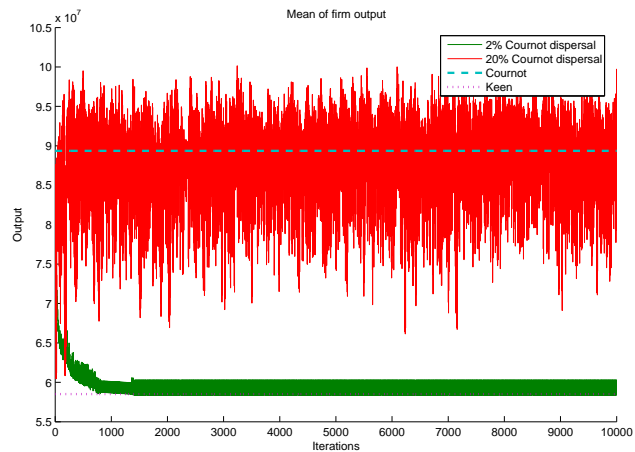}%
\caption{Comparative stability of average firm for different values of $\delta
q$}%
\label{Comparison_dq_Firms}%
\end{center}
\end{figure}

However, as $\sigma$ rises from 1\% to 20\% of the Cournot prediction, the
outcome shifts from the Keen to the Cournot level (Figure
\ref{DispersalRisingMC}). Aggregate and individual agent behavior also becomes
much more unstable, as Figure \ref{Comparison_dq_Firms} indicates. We surmise
that the \textquotedblleft emergent collusion\textquotedblright\ we identified
in \cite{StandishKeen}\ breaks down as $\delta q$ rises; perhaps the
increasing size of unpredictable output changes by other firms makes the
overall market environment more chaotic, forcing each firm to rely more on
feedback from its own output.changes.%

%

\section{Conclusion}

Contrary to the beliefs of the vast majority of economists, equating marginal
revenue and marginal cost is not profit-maximizing behavior, the number of
firms in an industry has no discernible impact on the quantity produced, the
\textquotedblleft deadweight loss of welfare\textquotedblright\ exists
regardless of how many firms there are in the industry, and instrumentally
rational profit-maximizers do not play Cournot-Nash games. Moving from
Hollywood to The Bard, it appears that the dominant theory of the firm is
\textquotedblleft Much Ado About Nothing\textquotedblright.

\end{document}